# Unidirectional domain growth of hexagonal boron nitride thin films

Abhijit Biswas,[1,*] Qiyuan Ruan,[1] Frank Lee,[2] Chenxi Li,[1] Sathvik Ajay Iyengar,[1] Anand B. Puthirath,[1] Xiang Zhang,[1] Harikishan Kannan,[1] Tia Gray,[1] A. Glen Birdwell,[3] Mahesh R. Neupane,[3] Pankaj B. Shah,[3] Dmitry A. Ruzmetov,[3] Tony G. Ivanov,[3] Robert Vajtai,[1] Manoj Tripathi,[2,*] Alan Dalton,[2] Boris I. Yakobson[1,4,*] and Pulickel M. Ajayan[1,a*]

**AFFILIATIONS**

[1]Department of Materials Science and Nanoengineering, Rice University, Houston, Texas 77005, USA

[2]Department of Physics and Astronomy, University of Sussex, Brighton BN1 9RH, United Kingdom

[3]DEVCOM Army Research Laboratory, RF Devices and Circuits, Adelphi, Maryland 20783, USA

[4]Department of Chemistry, Rice University, Houston, Texas 77005, USA

[a)]**Authors to whom correspondence should be addressed:** abhijit.biswas@rice.edu, m.tripathi@sussex.ac.uk, biy@rice.edu, ajayan@rice.edu

**Abstract**

Two-dimensional van der Waals (2D-vdW) layered hexagonal boron nitride (h-BN) has gained tremendous research interest over recent years due to its unconventional domain growth morphology, fascinating properties and application potentials as an excellent dielectric layer for 2D-based nano-electronics. However, the unidirectional domain growth of h-BN thin films directly on insulating substrates remains significantly challenging because of high-bonding anisotropicity and complex growth kinetics than the conventional thin films growth, thus resulting in the formation of randomly oriented domains morphology, and hindering its usefulness in integrated nano-devices. Here, ultra-wide bandgap h-BN thin films are grown *directly* on low-miscut atomically smooth highly insulating *c*-plane sapphire substrates (*without using any metal catalytic layer*) by pulsed laser deposition, showing remarkable unidirectional triangular-shape domains morphology. This unidirectional domain growth is attributed to the step-edge guided nucleation caused by reducing the film-substrate interfacial symmetry and energy, thereby breaking the degeneracy of nucleation sites of random domains, as revealed by the density functional theory (DFT) calculations. Through extensive characterizations, we further demonstrate the excellent single crystal-like functional properties of films. Our findings might pave the way for feasible large-area direct growth of electronic-quality h-BN thin films on insulating substrates for high-performance 2D-electronics, and in addition would be beneficial for hetero engineering of 2D-vdW materials with emergent phenomena.

# Graphical abstract

We demonstrated the unidirectional domain growth of h-BN directly on insulating sapphire by pulsed laser deposition method, attributed with the step-edge guided nucleation caused by reducing the film-substrate interfacial symmetry. The step-edge-guided aligned growth might pave the way for large-area synthesis of electronic-quality epitaxial h-BN and consequent emergent phenomena.

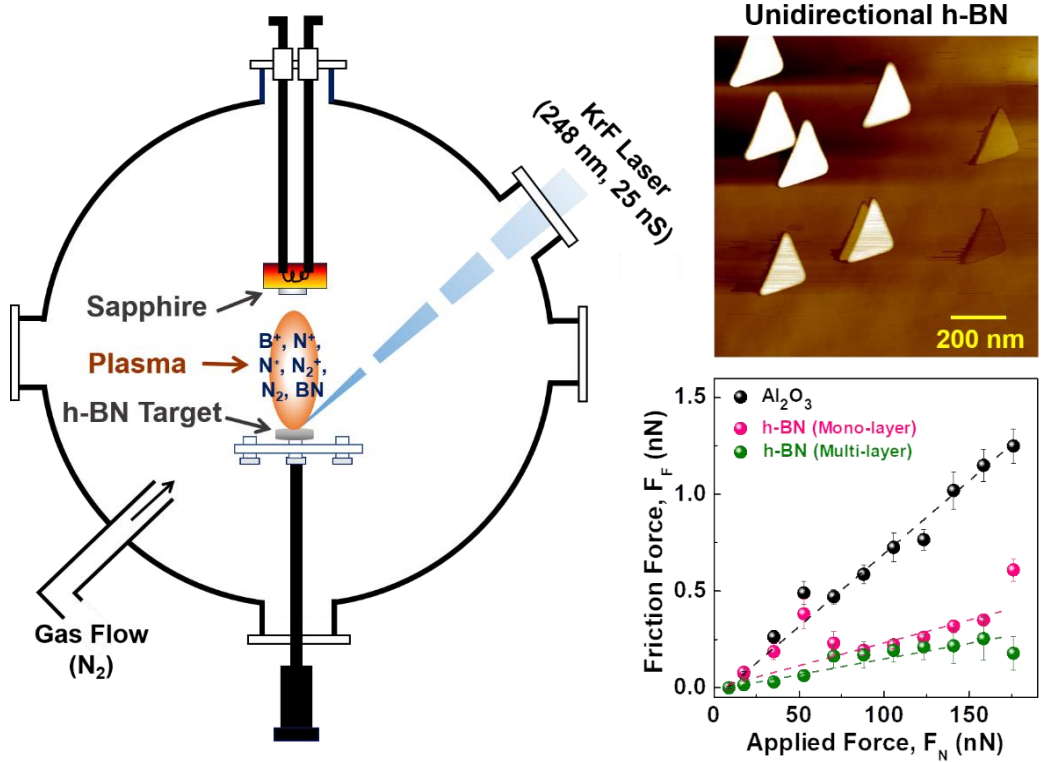

# Highlights

**[1].** We demonstrated unidirectional domain growth of 2D-vdW h-BN thin films directly on atomically smooth low miscut insulating sapphire substrates.

**[2].** The unidirectional domain growth is attributed to the step-edge guided nucleation caused by reducing the film-substrate interfacial symmetry, as revealed by the DFT calculations.

**[3].** Through extensive characterizations, we show the excellent single crystal-like functional properties.

**[4].** It is important for large-area direct growth of electronic-quality h-BN for high-performance 2D-electronics, and would be beneficial for hetero engineering of 2D-materials.

**Introduction**

The growth of layered two-dimensional van der Waals (2D-vdW) materials has been a topic of tremendous interest not only for the fundamental understanding of underlying growth mechanisms because of their high-bonding anisotropicity but also for diverse functionalities that are useful for several technological applications[1-3]. Interestingly, a majority of 2D-vdW materials form layered hexagonal structures and it is evident from their microscopic and optical images that they show triangular-shape domains morphology with lateral dimensions ranging from a few hundred nanometers to several micrometers, and research efforts to grow large-area high-quality thin films of these materials is still ongoing [4]. Nevertheless, most of the research has shown grown multi-domains are aligned to some extent, often not unidirectionally, but also not randomly over substrate surfaces, causing defects at grain boundaries [3,5]. This is argued to be related with the film-substrate interaction energy along a specific direction, film growth kinetics, and the surface quality of the substrates. Recently, several groups have reported that substrate miscut angle, orientations, and consequent step-terrace plays an important role in the formation of unidirectional triangular-shape domains with different lateral sizes for various 2D-vdW films, grown by metal-organic chemical vapor deposition (MOCVD) [6-8]. It is attributed to the interplay between the film-substrate symmetry at the step-edge, which breaks the degeneracy of nucleation energy for antiparallel domains and leads to the unidirectional alignment of domains [6-13].

Considering the huge technological importance [14] of 2D-vdW hexagonal boron nitride (h-BN) layers, extensive efforts have been made to grow h-BN thin films by MOCVD [15-20]. h-BN is a unique 2D-vdW material exhibiting excellent structural, chemical, thermal, optical, electrical, and tribological properties,[21,22] and is considered to be the best substrate for the graphene growth [23]. Given the potential advantages for 2D-based nano-electronics owing to its excellent chemical and thermal stability, wide-bandgap ($E_g$ ~6 eV), and low dielectric constant ($\varepsilon = 5$), h-BN is one of the most promising gate insulators [14]. Recently, unidirectional domains growth of h-BN was also demonstrated, by utilizing metal (Cu) as catalytic support, thereby minimizing the interaction energy on the inclined surface of the catalyst (strong coupling between the h-BN and metal surface) [17-20]. There are several reports also exists about the mis-aligned BN domains growth on metal surfaces [24-28]. However, growth of h-BN unidirectional domains directly on insulating substrates, without using any catalytic support, is important to understand the growth dynamics, morphology, and surface termination. More importantly, it is beneficial for wafer-scale film growth and as a smooth interfacial thin dielectric layer, eliminating the grain-boundaries, and reducing the substrate-induced or passivation layer induced degradation of the carrier mobility of 2D-vdW materials [**29**]. Furthermore, the unidirectional aligned growth of h-BN domains would be beneficial as a template to engineer 2D/2D and 2D/3D hybrid heterostructures, wide spreading the horizon of novel 2D-vdW materials research, with the observation of unprecedented emergent functionalities [**30**].

In the present work, we have successfully grown h-BN thin films *directly* onto low-miscut ($\leq 0.1°$) atomically smooth insulating *c*-Al$_2$O$_3$ (0001) sapphire substrates, by using a highly energetic thin film deposition technique called pulsed laser deposition (PLD). Remarkably, h-BN films show unidirectional domains growth morphology without using any catalytic support, unlike previous reports [**17-20**]. Structural, morphological, spectroscopic, optical, magnetic, and tribological characterizations confirm the growth of h-BN films with characteristic properties. We demonstrate that possibly growth kinetics, a low-miscut of the substrate along a specific direction and its atomically flat surface reduce the film-substrate symmetry and interfacial interaction energy, thereby promoting the nucleation along the step-edges and resulting in the formation of unidirectional domain growth morphology, revealed by theoretical calculations.

**Results and discussion**

*Crystalline quality of c-Al$_2$O$_3$ (0001) sapphire substrate for film growth*

For the growth of h-BN, we chose hexagonal *c*-plane sapphire (*c*-Al$_2$O$_3$) substrate as it is commonly used for 2D-materials growth. It is highly insulating ($E_g$ ~7 eV), non-magnetic, and most importantly structurally matched with h-BN (both having three-fold structural symmetry). The in-plane lattice parameters are ~4.78 Å (*c*-Al$_2$O$_3$) and ~2.50 Å (h-BN). From the epitaxy relationship, they show ~1:2 commensurate lattice matching ($a_{\text{c-Al2O3 (0001)}} \approx 2a_{\text{h-BN (0001)}}$) with ~4.4% tensile strain (calculated as $\left(\frac{a_s - a_f}{a_f}\right) \times 100\ \%$ where $a_s$ is the substrate in-plane lattice constant and $a_f$ is the film lattice constant) onto h-BN. As substrate-film interfacial interaction plays a crucial role in the growth morphology of 2D-vdW materials [**9**], thus we first characterized the substrate quality by measuring its miscut angle ($\alpha$), as non-zero miscut is unavoidable and limited by the accuracy of the cutting tools. A typical as-received substrate is shown along with the cut-directions i.e. along A-plane (11-20) [**Fig. 1a**]. Thin-film X-ray diffraction (XRD) shows the single-crystalline quality of the substrate [**Fig. 1b**]. Importantly, we measured the miscut angles along the A-plane as well as the M-plane (10-10) by measuring the rocking curve of the (0006) Bragg peak at four different azimuthal ($\varphi$) angles [**Fig. 1c**]. The miscut angles were obtained as $\alpha_{C/A}$ ~0.089° and $\alpha_{M/A}$ ~0.195°, where C/A is the miscut towards A-axis and C/M is the miscut towards M-axis, respectively [**Fig. 1c**]. The full-width at half maxima (FWHM) of the rocking curve was also found to be ~0.002°, confirming excellent crystalline quality. Moreover, atomic force microscopy (AFM) topographic images show the smooth surface of as-received and high-temperature annealed substrate with a step-terrace structure having a low surface roughness of ~0.223 nm and ~0.157 nm [**Figs. 1d** and **1e**].

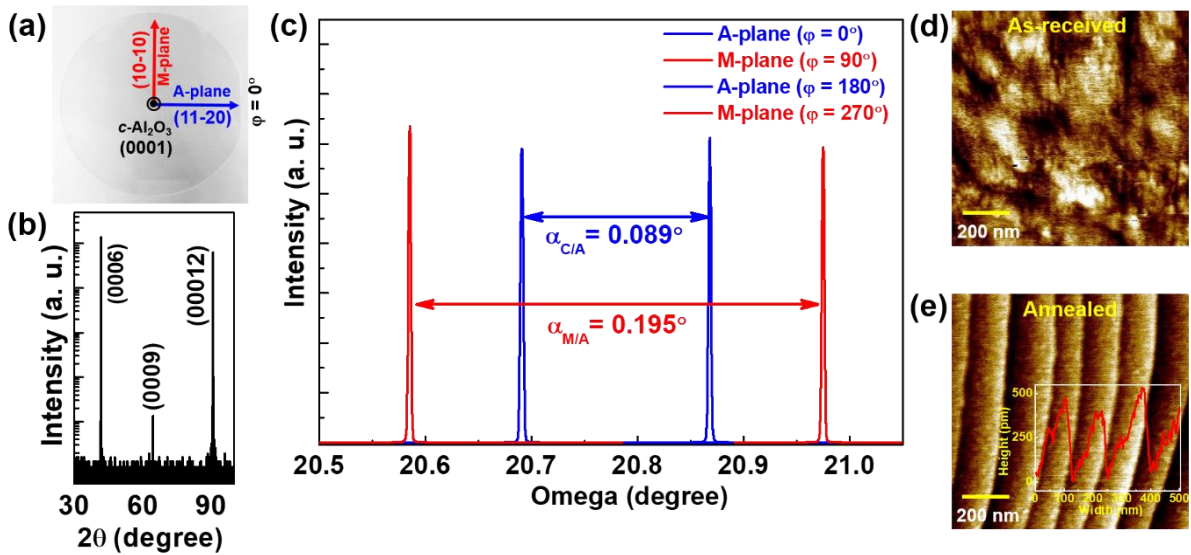

**FIGURE 1**. **Characterizations of *c*-Al$_2$O$_3$ (0001) sapphire substrate**. (a) As-received 2-inch substrate, showing the crystal planes and mechanical cut-direction. (b) X-ray diffraction pattern shows the crystallinity. (c) Omega scans along with different azimuthal angles (both along with A and M-plane), shows the miscut angles ($\alpha$) with $\alpha_{C/A}$ ~0.089° and $\alpha_{M/A}$ ~0.195°. Atomic force microscopy images of an as-received substrate (d) with roughness ~0.223 nm, and a high-temperature annealed atomically flat surface (e) with roughness ~0.157 nm.

*Chemical and spectroscopic characterizations of films*

We grew various thicknesses of BN films by using PLD, a highly energetic thermally non-equilibrium process with a true stoichiometric transfer of target elements onto the substrate in the form of plasma, with enhanced adsorbate surface mobility endowed by the high energy (~10-100 eV) radicals (both ionized and neutral) [31]. After the growth, we characterized films by using X-ray photoelectron spectroscopy (XPS), which shows the bonding characteristics between the elements. As seen, all the films show the formation of characteristic B-N bonds with peak centers at ~190.8 eV (B 1s core) and at ~398.2 eV (N 1s core), respectively [**Figs. 2a, 2b** and supplementary material **Fig. S1**] [**32-35**]. Moreover, from XPS elemental scans the B:N ratios were found to be ~ 1:1. To confirm the h-BN phase formation and bonding structures, we performed Fourier-transform infrared spectroscopy (FTIR), which shows the characteristic intrinsic $E_{2g}$ h-BN peak at ~1360 cm$^{-1}$ [**Fig. 2c**], corresponding to the transverse optical (TO) vibrations for in-plane B-N bond stretching in sp$^2$-bonded h-BN [**36**]. Furthermore, we also performed the Raman spectroscopy, which also shows the sharp $E_{2g}$ in-plane Raman peak at ~1360±4 cm$^{-1}$ [**Fig. 2d** and supplementary material **Fig. S2**]. The FWHM of the Raman peak was found to be ~14.06 cm$^{-1}$ [inset of **Fig. 2d** and supplementary material **Fig. S1**], and is slightly higher than the observed FWHM of ~8 cm$^{-1}$ for bulk ordered single crystal h-BN [**37**]. These observations confirm the stoichiometric high-quality growth of h-BN thin films on low-miscut atomically smooth sapphire substrates by using PLD.

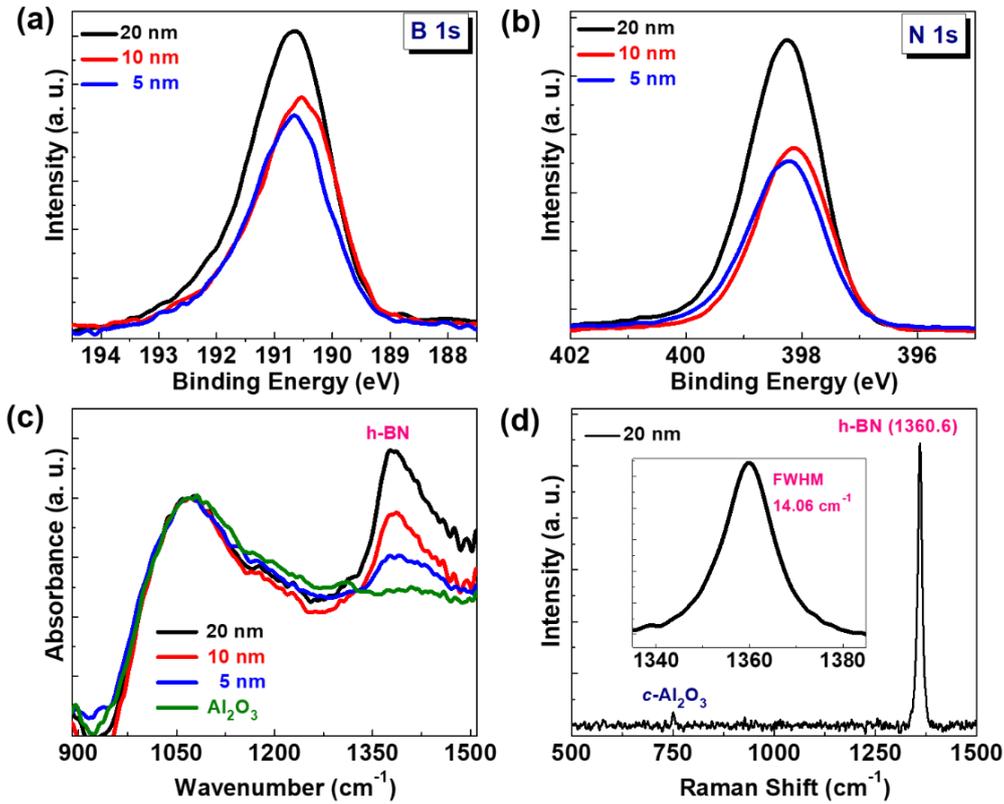

**FIGURE 2**. **Thickness-dependent structural characterziation of h-BN films**. XPS at (a) B 1s and (b) N 1s core confirm B-N bonds. (c) FTIR spectra show the sp$^2$ bonded in-plane vibrational transverse optical (TO) mode of h-BN. (d) Raman spectra of a ~20 nm film show the characteristic $E_{2g}$ Raman peak at ~1360 cm$^{-1}$ for h-BN. Inset shows the full-width at half maxima of the observed Raman peak.

***Topography of h-BN films***

We obtained the surface morphology of various thicknesses of h-BN films by using atomic force microscopy (AFM). Interestingly, we observed typical triangular-shape geometry (characteristics of three-fold symmetric h-BN) [**Fig. 3** and supplementary material **Fig. S3**] for all films. Here to be mentioned that to the best of our knowledge, all the existing reports show the multidirectional 3D-island-type surface morphology for h-BN films grown by PLD [**38-43**]. Remarkably, for our films these triangular domains are oriented unidirectional rather than randomly (also see the optical image in supplementary material **Fig. S4**). Until now, reported similar unidirectional domain growth directly on a sapphire was limited to only MoS$_2$, WS$_2$ and WSe$_2$ films [**6-8**]. The development of unidirectional geometry relates to the initial growth at the step-edges with minimal substrate-film interaction energy and lower surface symmetry along a specific plane, thus breaking the degeneracy of nucleation energy for the antiparallel domains, and preferring unidirectional alignment of domains, as illustrated through the schematics [**Fig. 3d**]. Similarly, Li *et al*. demonstrated the unidirectional alignment depends on the step orientations (i.e., C/A or C/M), and C/A shows the formation of unidirectional domains, whereas C/M forms two anti-parallel domains crossing the step edges [**6**]. It is known that high-temperature annealing reconstructs the sapphire surface and forms step-terrace structures, which has a significant influence on the aligned growth of graphene [**11**].

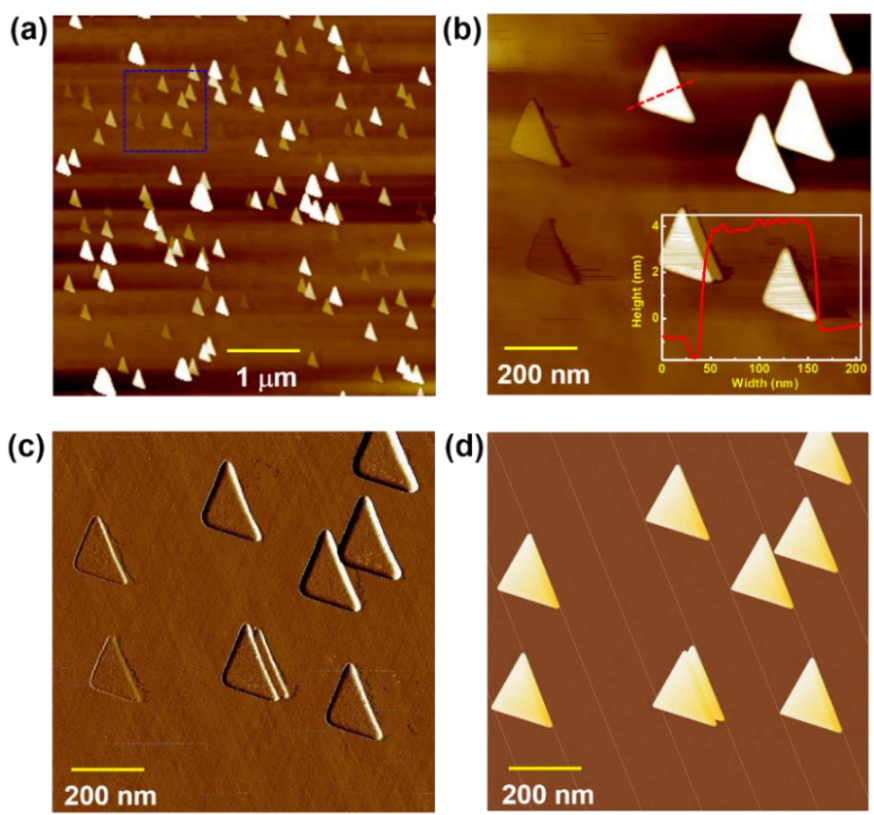

**FIGURE 3**. **Surface topography of h-BN film**. (a) The unidirectional triangular-shape domains morphology. (b), (c) Topography of the square blue dotted portion in panel (a) and its corresponding phase image. (d) Schematic illustrating the unidirectional aligned domains at the step edges of the sapphire.

For our h-BN growth, *c*-plane sapphire substrates with a very low-miscut angle ($\alpha_{C/A}$ ~0.089°) were used. Before the depositions, we also performed in-situ pre-annealing of the sapphire substrates at the same deposition temperature in order to reconstruct the surface further, as well as remove the surface contaminants. Thus, in addition to the high-energy growth kinetics of a PLD process and a very low-miscut angle of the used sapphire ($\alpha_{C/A}$ ~0.089°), this pre-annealing also helps in forming smoother perfect step-terrace structures. This reduced the step-edge symmetry of the substrate and made it possible to have all the h-BN domains aligned only along a particular direction, thereby forming an oriented growth along the step-edges [**9**]. The lateral sizes of these triangular-shaped unidirectional domains are ~100-250 nm [**Fig. 3b** and supplementary material **Fig. S3** and **S5**]. These are much smaller than the h-BN triangular domains grown by CVD (~few μm) [**17-20**], and may be self-limited due to the growth conditions used and related growth kinetics during the deposition. In addition, the density as well as the step-terrace of each step of the substrate will also affect the nucleation process, however, mainly on the density of nuclei. Denser step-edges will carry more nuclei leading to more islands, although the lateral sizes of these domains will still be determined by the terrace width of the substrate surface. At present, it is difficult to comment on the actual crystalline nature of these domains, without the in-depth nanoscale atomically resolved images. However, the FWHM of $E_{2g}$ Raman peak is lower compared to the previously reported h-BN thin films grown by CVD or PLD (supplementary material **Table S1**), suggesting possible high crystalline quality of these domains. Nevertheless, considering the rapid and clean growth process of PLD and its rapid advancement in understanding the in-situ growth evolution, further efforts are needed to grow large-area continuous single-crystalline electronic-quality h-BN thin films.

*Early-stage growth evolution and step-edge guided growth of films*

The unidirectional nucleation of h-BN triangles during the early stage of growth is also shown in **Fig. 4**. The topography and especially the phase imaging presents early-stage growth (and growth evolution) of h-BN films [**Figs. 4a**, **4b** and supplementary material **Fig. S6**], and is marked by the dashed yellow regions depicting that growth begins with the formation of triangular-shape domains that are align unidirectional. This is attributed to the growth kinetics at the step-edges as the phase image of a film clearly shows the distribution of the h-BN domains mostly over the step-edges [**Fig. 4c** and supplementary material **Fig. S7**] [**44**] and signifies the importance of the highly energetic non-equilibrium growth process of PLD.

A closer inspection reveals that rather than the lateral increase in domain sizes with extended growth time, continued deposition takes place either on the top of the existing domains or as new domains at the step-edges. Remarkably, they are also oriented unidirectional. The growth process is shown by the schematic illustrations [**Fig. 4d**]. With the underlying anisotropic h-BN film growth rate,[**19**] for thicker film growth the substrate surface eventually gets fully covered, consistent with the observed thickness-dependent morphology [supplementary material **Fig. S3**]. This indeed confirms that the highly symmetric step-edge of the substrate can serve as nucleation sites to initiate the unidirectional growth of h-BN, rather than the randomly oriented growth and with the growth evolution, both vdW epitaxy and step-edge guided nucleation contributed to the aligned growth.

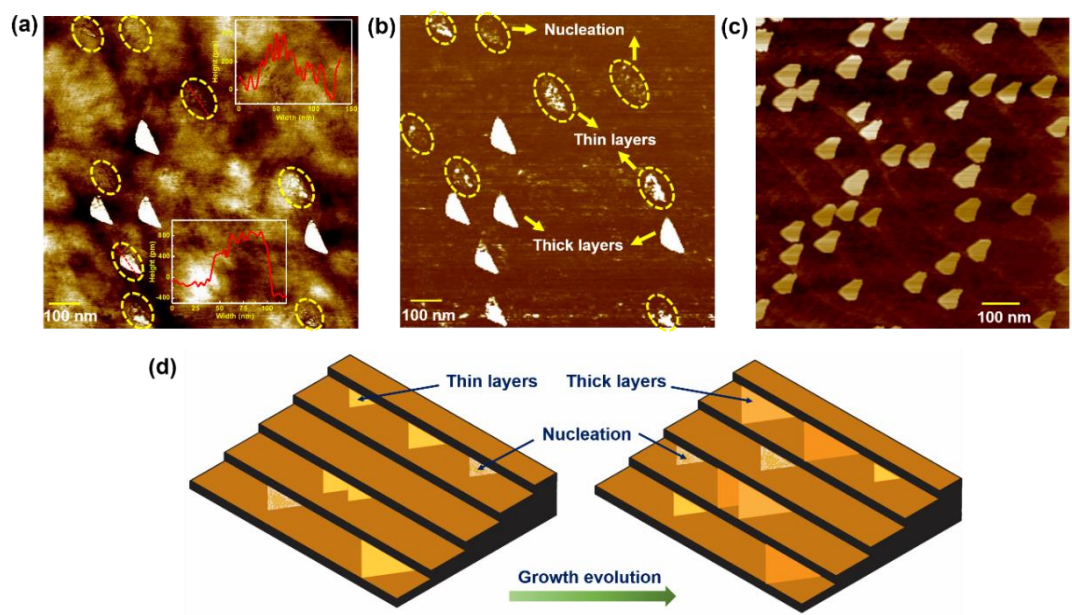

**FIGURE 4. Evolution of unidirectional triangular-shape domains and step-edge guided growth of h-BN film**. (a) Surface topography (inset shows the height profiles for ultra-thin domains) and corresponding (b) phase image of ultrathin domains growth evolution. (c) Clearly visible step-edge guided unidirectional domain growth, and (d) schematic illustrations show the nucleation, thin layers, and thick layers of unidirectional domains, rather than random orientation during the growth.

*Insights into the mechanism from density functional theory calculations*

To understand the cause of the unique oriented growth nature of h-BN directly on sapphire, we evaluate the nucleation structures and energies involved by using the density functional theory (DFT) calculations. Recent quantitative studies suggested, for other systems, especially on metal substrates, that the oriented growth morphology is controlled not by plane-to-plane van der Waals interaction (which is too weak for initially-small nuclei), but rather by much stronger attachment of h-BN edge to the step of substrate, leading to probable unidirectional domains.[**9,13**] In order to find out if there is an interfacial orientation preference, we evaluate the step-edge binding during the nucleation process.

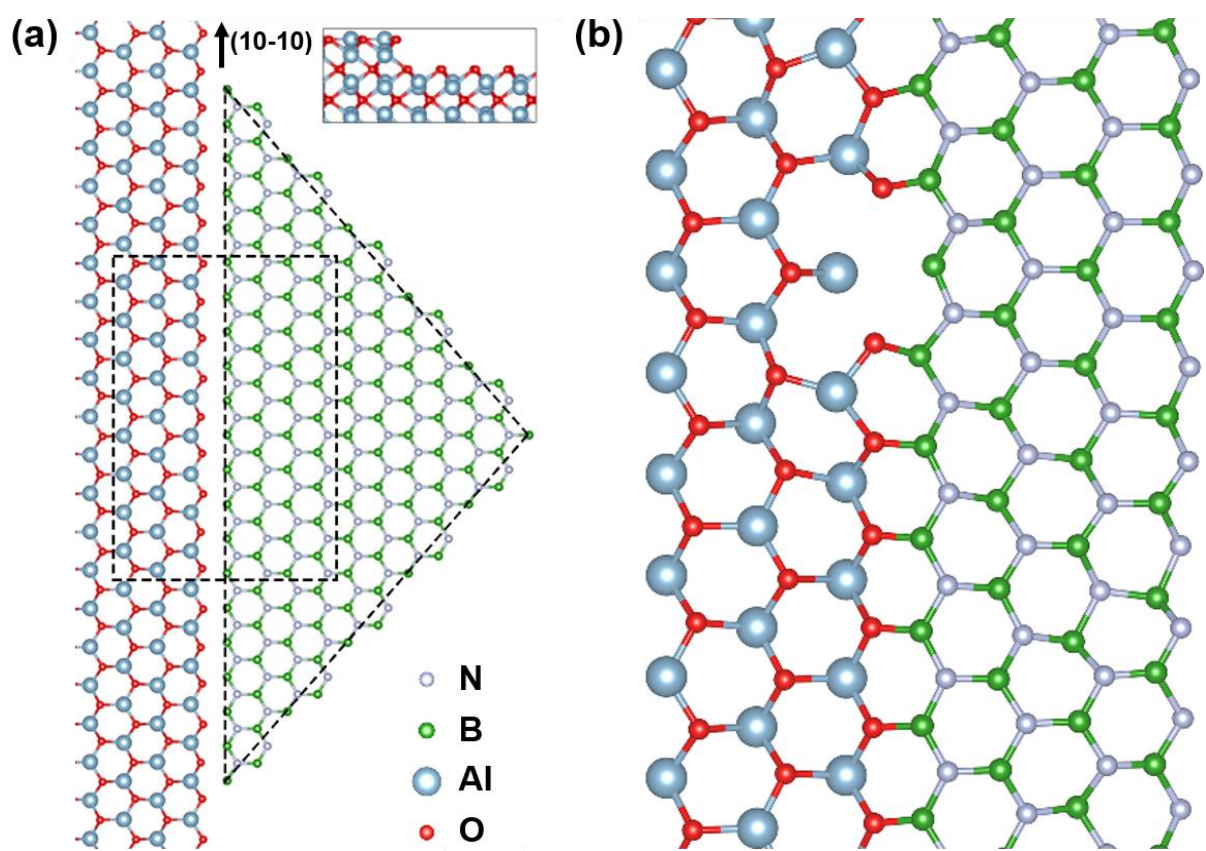

**FIGURE 5. Schematics of zigzag h-BN/*c*-Al$_2$O$_3$ interface.** (a) A top view illustration of the interface between O-terminated sapphire step-edge along (10-10) direction and B-terminated ZZ h-BN edge. The inset shows the side view of the step-edge configuration in the substrate. The dashed triangle is the seed of the triangular domain in experiments. (b) The relaxed structure of the dashed rectangle in (a). The defect corresponds to the oxygen atoms leaving the plane and moves down close to the second Al$_2$O$_3$ layer, however still bonded with the boron atom. For clarity, only the Al/O atoms in the topmost layer are shown here.

For the calculations, O-terminated step-edge along (10-10) direction [inset in **Fig. 5a**] of $Al_2O_3$ and the zigzag (ZZ) edge of h-BN are placed in contact [**Fig. 5a**], as ZZ edge of h-BN always shows stronger binding to substrates,[19] as well smaller lattice mismatch than the armchair (AC) h-BN with O-terminated $Al_2O_3$. The lattice parameter mismatch of 11% is compensated by a misfit edge dislocation (Burgers vector |b| = 2.5 Å) appearing spontaneously after structural optimization, **Fig. 5b**; their macroscopic density is accordingly ~$4\times10^6$ cm$^{-1}$. In contrast, if another stable edge of h-BN, the AC one was attached to the step-edge, its B/N alternate atomic configuration will never lead to two same side edges [supplementary material **Fig. S8a**], which means the two base angles of the triangular-island will be different. In addition, swapping the B/N atoms in AC edge [supplementary material **Fig. S8b**] does not change the interface energy and thus the AC contact would lead to the equiprobable formation of two inverted types of islands at the step-edges. In our experiments, all the triangles have identical shapes with no distinct differences, which rules out the possibility of h-BN AC edge. The DFT calculations show that, different from metal-substrate steps [20], the B-terminated ZZ edge has 1.0 eV/Å stronger binding to O-$Al_2O_3$ step than N-terminated ZZ [supplementary material **Fig. S9**]. This large energy preference should lead to a sole dominant interface configuration, ensuring eventually a unidirectional h-BN growth [**Fig. 5a**]. Furthermore, once the unidirectional domains are formed, the growth of the layers beyond the first monolayer is possibly controlled by the vdW interaction, and the h-BN stacking with lowest formation energies becomes dominant. The common h-BN stacking with the lowest energies are AA, AB, and AA' [13, 45]. As shown, AA' and AB stacking have the lowest total energies, in which AA' has a 60°-rotation angle between the first and second layer, whereas AB has 0°-rotation angle, maintaining the unidirectional orientation even for multilayer h-BN domains. Therefore, for our case, multilayer h-BN domains also possibly forms ABAB…stacking. This step-edge-guided aligned growth seems to be a universal feature, not limited to metal substrates, and can be generalized for other 2D-vdW materials, as we show recently for borophene [46].

*Bandgap and magnetic characterization of films*

Regarding the functional properties, e.g. bandgap, room-temperature UV-visible absorption spectrum of h-BN film showed a sharp peak at ~214 nm in the absorption spectrum [inset of **Fig. 6a**]. From the Tauc-plot, the direct bandgap was estimated to be $E_g$ ~5.95 eV [**Fig. 6a**], and is consistent with the bulk h-BN bandgap [22]. The measured bandgap also suggests that grown h-BN films are highly transparent [inset of **Fig. 6a**] and as well as highly insulating.

Through XPS valence band spectrum (VBS), we observed the band structure of the h-BN film with two distinct features with maxima at ~12 eV, and at ~20 eV, respectively [**Fig. 6b**] [**47,48**]. The valence band maxima (VBM) was at ~1.9 eV, from the Fermi level ($E_F$).

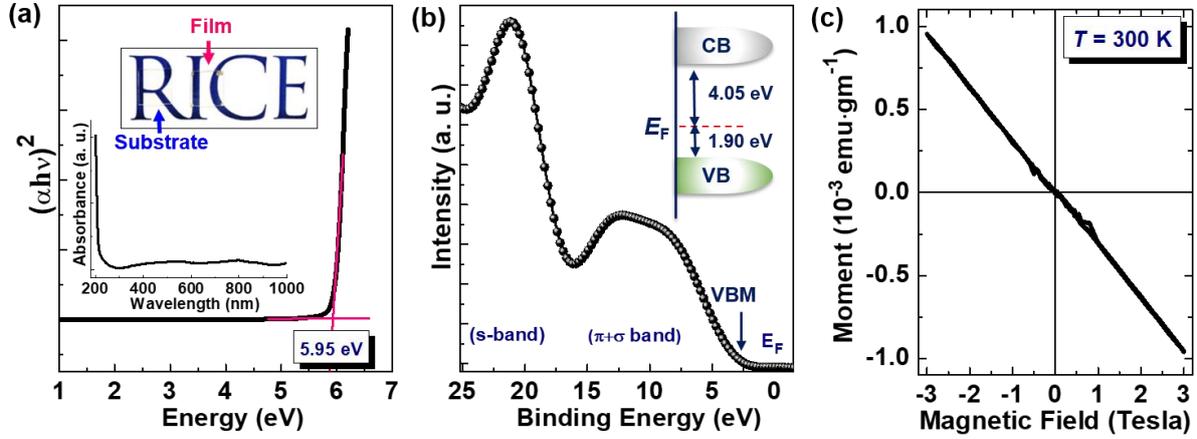

**FIGURE 6**. **Bandgap and magnetic characterization of h-BN films**. Room temperature (a) UV-visible absorbance spectra and corresponding direct-bandgap Tauc plot. Inset shows the images of a transparent h-BN film and sapphire. (b) Valence band spectrum of a film with valence band maxima (VBM) at ~1.9 eV. Inset shows the band-alignment. (c) Magnetization measurements demonstrating non-magnetic nature of h-BN film.

Additionally, we measured the magnetization of h-BN film. It is well-known that the perfect crystalline defect-free h-BN is a typical non-magnetic 2D-vdW material as it has both non-magnetic light-elements (B and N) without any unpaired electrons [**49**]. However, defective or functionalized (e.g. vacancies, adatoms, dopants) h-BN exhibits robust room-temperature ferromagnetism, due to the formation of local moments [**49**]. In our case, room-temperature magnetic hysteresis (M-H) measurement shows a perfect diamagnetic response [**Fig. 6c**], signifying that the grown h-BN films are almost free of defects or impurities, within the detection limit of the SQUID magnetic measurement system.

*Adhesion and friction force characterizations of films*

The tribological properties of h-BN film are characterized through adhesion force map and friction force microscopy (FFM) using a silicon cantilever (covered by native oxide) [**Fig. 7**]. The flexural deflection of the cantilever moving away from the scanned surface is measured as work of adhesion (pull-out force) and the torsional deflection due to lateral sliding results in friction force ($F_F$) values [**50**]. There is a significant contrast in adhesion force between h-BN and $c$-$Al_2O_3$ sub-surface, under similar load conditions [**Figs. 7a** and **7b**] revealing substantial contrast in the surface chemistry between the layer of $c$-$Al_2O_3$ and nearly inert film of h-BN. Nevertheless, the adhesion force values are very similar for each layer thickness from one-layer (1L) to multi-layers (ML) h-BN (~14 nm) [**Fig. 7d** and supplementary material **Fig. S10**]. The lower adhesion force in the h-BN sheet influences the $F_F$ response that drops significantly compared to $c$-$Al_2O_3$ [**Figs. 7b** and **7c**]. In contrast to adhesion force, the $F_F$ values vary with the thickness due to different shear resistance from each layer [**51**].

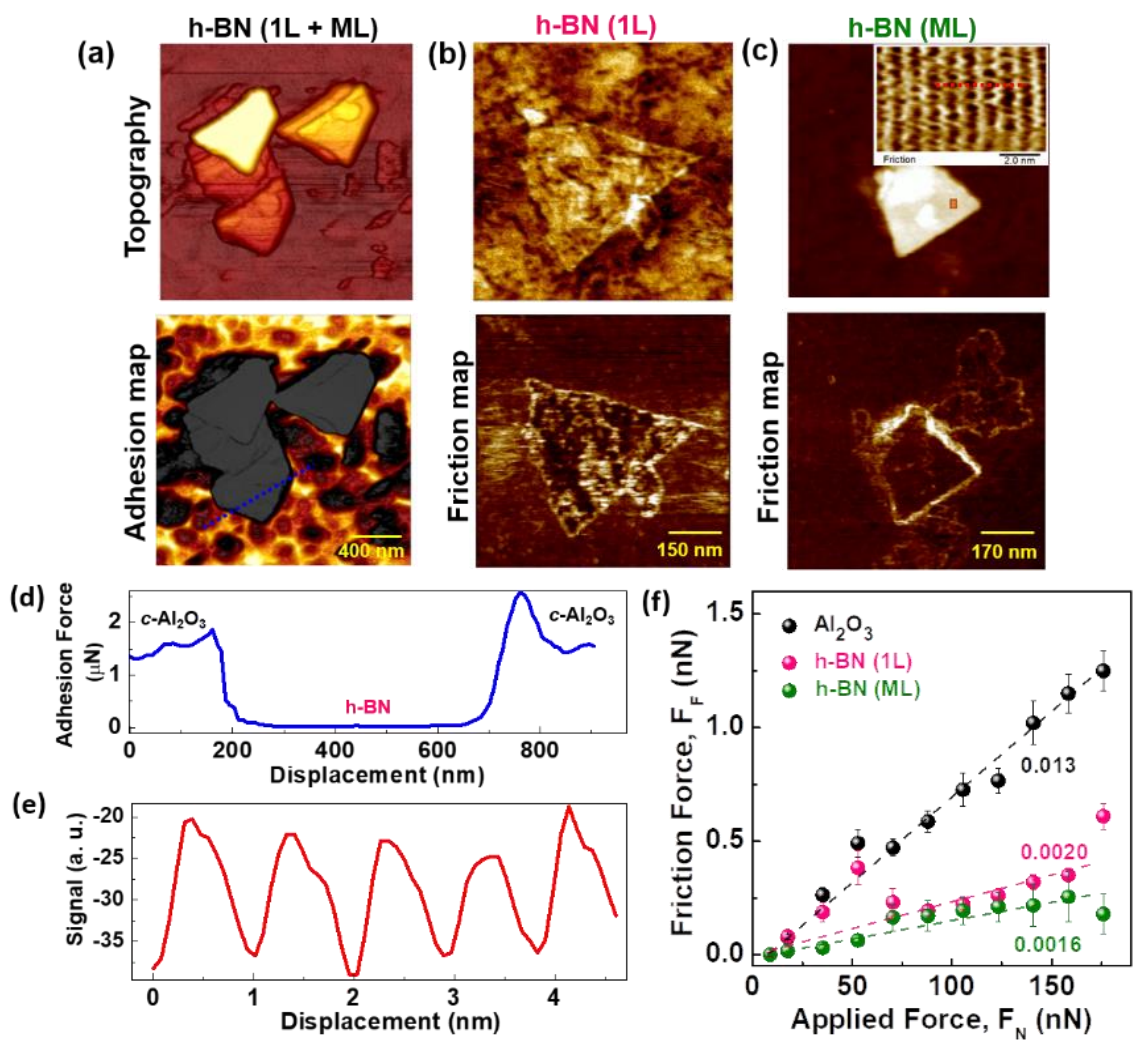

**FIGURE 7**. **Tribological characterization of h-BN domains**. (a) 3D topography and its concurrent adhesion force map of h-BN films over $Al_2O_3$ substrate. (b), (c) Topography of 1L h-BN and ML h-BN, and its concurrent friction force map, respectively. Inset in figure (c) illustrates the high-resolution friction map over an ML h-BN marked by the red rectangle. (d) The adhesion force (blue dotted line profile in Fig. **a**) reveals a significantly lower adhesion force of h-BN than the $Al_2O_3$ subsurface. (e) Atomic-scale stick-slip profile of the ML h-BN marked in panel (c) (red dotted line). (f) Load-dependent friction over $c$-$Al_2O_3$ substrate, 1L, and ML h-BN.

The atomic-scale visualization reveals the asymmetric stick-slip responses and unleashes the distinctive contribution from B and N atoms in the h-BN layer toward friction values [**Figs. 7c** and **7e**]. The load-dependent friction under applied force ($F_N$) ranging between ~8-175 nN [**Fig. 7f**] reveals friction response in the following order: $c$-$Al_2O_3$ > 1L h-BN > ML h-BN with their respective coefficient of friction (COF) measured through linear fit as ~0.013, ~0.0020, and ~0.0016. Moreover, monolayer h-BN shows higher $F_F$ value than its multilayer counterpart due to the presence of edges that are curvature induced due to the morphology of $c$-$Al_2O_3$. This could also explain the uneven distribution of the $F_F$ in 1L h-BN [**Fig. 7b**]. On the other hand, the ML h-BN is smoother and leads to a further drop in the $F_F$ values; thus promoting higher lubricity driven by the interlayer vdW forces. The tribological results indicate that PLD grown unidirectional h-BN films could be useful as an aligned lubricating surface.

**Conclusions**

In summary, we demonstrated unidirectional domain growth of 2D-vdW h-BN thin films directly on low-miscut atomically smooth sapphire substrates by PLD without using any metal catalytic support. Density functional theory calculation provides the insights into the mechanism of strong interplay between the film-substrate interfacial symmetry and energy, and the passivation of the nucleation sites along the preferred step-edge orientation of the substrate, responsible for the unidirectional domain growth. Several characterization techniques including chemical, spectroscopic, microscopic, optical, magnetic, and tribology confirms the growth of high-quality h-BN films exhibiting excellent intrinsic properties. Considering the importance of an ultrathin dielectric h-BN layer, our finding could pave the way for large-area unidirectional domain growth of electronic-quality thin films directly on insulating substrates for high-performance 2D-electronics. Moreover, this oriented domain growth could be extremely useful as a template to create "designer" 2D/2D or 2D/3D heterostructures displaying unforeseeable functionalities.

**Methods**

*Thin film growth (Pulsed laser deposition)*

Hexagonal boron nitride (h-BN) thin films were grown by PLD (KrF excimer laser with wavelength of 248 nm, and pulse width 25 nS). Films were grown by using the following deposition conditions: growth temperature ~800 °C, laser fluency ~2.2 J/cm$^2$ (laser energy ~230 mJ, spot size~ 7 mm×1.5 mm), repetition rate 5 Hz, target to substrate distance ~50 mm, and high-purity (5N) nitrogen gas partial pressure ($P_{N2}$) ~100 mTorr (flow rate ~73 sccm). Nitrogen was supplied to compensate the loss of nitrogen, if any and preserve the stoichiometry of BN ~1:1. We used commercially available high-purity (99.9% metal basis) one-inch diameter polycrystalline h-BN target [supplementary material **Fig. S11**] for the laser ablation (American Element). For depositions, we used hexagonal *c*-Al$_2$O$_3$ (0001) substrates, purchased from University Wafer. The suppliers provided the substrate cut-direction. The atomically flat surface was obtained by annealing at 1200 °C for ~2 hrs. Before the growth, substrates were in-situ pre-annealed at the same growth temperature of 800 °C, for ~30 min. We mechanically cut the 2-inch substrate by diamond pencil and used ~4×4 mm$^2$ sizes for the growth. After growth, films were cooled down at ~20 °C/min for further characterizations. Typically, ~5 nm flakes were obtained by providing 500 laser shots (whereas 1000 laser shots for 10 nm, and 2000 shots for 20 nm). Since we used 5 Hz repetition rate (i.e. 5-laser pulses/sec) therefore the growth time was 100 sec for 5 nm BN and so on. For the early stage growth evolution, we provided 5/10/100 laser shots.

*Structural, chemical, and optical characterizations (XRD, XPS, VBS, FTIR, AFM, Raman, FESEM, Optical image and UV-visible)*

X-ray diffraction pattern was obtained with Rigaku SmartLab X-ray diffractometer, by using a monochromatic Cu K$_\alpha$ radiation source. X-ray photoelectron spectroscopy was performed by using PHI Quantera SXM scanning X-ray microprobe with 1486.6 eV monochromatic Al K$_\alpha$ X-ray source. High-resolution elemental scans and VBS were recorded at 26 eV pass energy. FTIR was obtained by using the Nicolet 380 FTIR spectrometer, equipped with a single-crystal diamond window. Park NX20 AFM was used to obtain surface topography, operating in tapping mode using Al-coated Multi75Al cantilevers. Raman spectroscopy measurements were performed using a Renishaw inVia confocal microscope. A 532 nm laser was used as the excitation source. Optical micrograph was obtained under bright field illumination at 100x magnification using a Labomed LB-616 Trinocular Microscope. UV-vis absorbance measurements were acquired using a Shimadzu 2450 UV-visible spectrophotometer. The topography of the target surface was performed by field emission scanning electron microscope (FE-SEM) (FEI Quanta 400 ESEM FEG).

*Magnetic characterization*

Room temperature magnetic hysteresis (M-H) measurement was performed using a SQUID-VSM-7 Tesla (Quantum Design, USA). Hysteresis loop was measured within the magnetic field sweep range of ±3 Tesla.

*Adhesion and friction force characterizations*

Atomic force microscopy (from Bruker Ltd.) has been operated in different modes for adhesion and friction force measurements as Peakforce tapping microscopy and lateral force microscopy (LFM), respectively. Silicon cantilever (diameter up to 10 nm) covered with native oxide has been used for the friction force investigations (Model: CSG10, $K_n$ =0.5±0.1, Resonance frequency = 35 kHz). The calibration of the cantilevers was carried out through two different methods of thermal tuning and Sader's method [**52**]. Under Peakforce operation, the adhesion force measurement is carried out through force-distance spectroscopy, where broad tip apex up to ~50 nm has been vertically approached to the surface at higher applied force (~90-100 nN) for greater contact area at the interface. The adhesion force is measured as a pullout force during the withdrawing action of the cantilever from the surface.

*Density functional theory (DFT) calculations*

DFT calculations were performed using the Perdew-Burke-Ernzerhof (PBE) generalized gradient approximation supported by the Vienna Ab-initio Simulation Package [**53,54**]. The plane-wave kinetic energy cutoff was set at 450 eV. A vacuum slab of 15 Å is used to avoid the periodicity in z-direction, and *a* ~25 Å interface is used to eliminate the lattice mismatch between substrate and h-BN. Atoms in h-BN and the top three layers of $Al_2O_3$ are relaxed until the force on each atom reaches 0.02 eV/Å. DFT-D3 vdW approach with Becke-Johnson damping and spin-polarization are used to catch the possible energy corrections [**55,56**].

**Supplementary materials**

See the supplementary material for the Raman spectra and the FWHM comparison with the reported literature, thickness dependent AFM and adhesion force of films and theoretical calculations, as well as the XRD of the bulk polycrystalline target used for the laser ablation.

**Acknowledgements**

This work was sponsored by the Army Research Office and was accomplished under Cooperative Agreement Number W911NF-19-2-0269. The views and conclusions contained in this document are those of the authors and should not be interpreted as representing the official policies, either expressed or implied, of the Army Research Office or the U.S. Government. The U.S. Government is authorized to reproduce and distribute reprints for Government purposes notwithstanding any copyright notation herein. We would like to thank Dr. Muhammad Rahman, Dr. Bin Gao, Dr. Bo Jiang, and Dr. Jianhua Li for their help in various measurements and discussions. Manoj Tripathi and Alan Dalton would like to thank strategic development funding from the University of Sussex.

**Conflict of Interest**

The authors declare that they have no known competing financial interests or personal relationships that could have appeared to influence the work reported in this paper.

**Author Contributions**

A. B., R.V. and P. M. A conceptualized the study. A.B., C. L., S. A. I., A. B. P., H. K., X. Z., and T. G. grew and characterized the films. F. L., M. T., and A. D. carried out the tribology measurement. Q. R. and B. I. Y. performed the density functional calculations. A. G. B., M. R. N., P. B. S, D. A. R. and T. I. commented on the data analysis. All the authors discussed the results and contributed on the manuscript preparation.

**Data Availability**

The datasets generated during and/or analyzed during the current study are available from the corresponding author on reasonable request.

**CRediT authorship contribution statement**

**Abhijit Biswas**: Conceptualization, Methodology, Validation, Formal analysis, Investigation, Data curation, Writing - original draft, Writing - review & editing, Visualization. **Qiyuan Ruan**: Formal analysis, Investigation, Data curation, Writing - original draft, Writing - review & editing, Visualization. **Frank Lee**: Methodology, Formal analysis, Investigation, Data curation, Writing - review & editing. **Chenxi Li**: Data curation, Formal analysis, Investigation. **Sathvik Ajay Iyengar**: Data curation, Formal analysis, Writing - review & editing. **Anand B. Puthirath**: Formal analysis, Data curation, Writing - review & editing. **Xiang Zhang**: Formal analysis, Data curation, Writing - review & editing. **Harikishan Kannan:** Formal analysis, Data curation, Writing - review & editing. **Tia Gray**: Formal analysis, Data curation, Writing - review & editing. **A. Glen Birdwell**: Formal analysis, Writing - review & editing. **Mahesh R. Neupane**: Formal analysis, Writing - review & editing. **Pankaj B. Shah**: Formal analysis, Writing - review & editing. **Dmitry A. Ruzmetov**: Formal analysis, Writing - review & editing. **Tony G. Ivanov**: Formal analysis, Writing - review & editing. **Robert Vajtai**: Conceptualization, Validation, Formal analysis, Supervision, Funding acquisition, Writing - original draft, Writing - review & editing. **Manoj Tripathi**: Methodology, Validation, Formal analysis, Investigation, Data curation, Writing - original draft, Writing - review & editing, Visualization. **Alan Dalton**: Supervision, Writing - review & editing. **Boris I. Yakobson**: Formal analysis, Investigation, Data curation, Writing - original draft, Writing - review & editing, Visualization. **Pulickel M. Ajayan**: Conceptualization, Validation, Supervision, Funding acquisition, Writing - original draft, Writing - review & editing.

# Supplementary Materials

# Unidirectional domain growth of hexagonal boron nitride thin films

Abhijit Biswas,[1,*] Qiyuan Ruan,[1] Frank Lee,[2] Chenxi Li,[1] Sathvik Ajay Iyengar,[1] Anand B. Puthirath,[1] Xiang Zhang,[1] Harikishan Kannan,[1] Tia Gray,[1] A. Glen Birdwell,[3] Mahesh R. Neupane,[3] Pankaj B. Shah,[3] Dmitry A. Ruzmetov,[3] Tony G. Ivanov,[3] Robert Vajtai,[1] Manoj Tripathi,[2,*] Alan Dalton,[2] Boris I. Yakobson[1,4,*] and Pulickel M. Ajayan[1,a*]

**AFFILIATIONS**

[1]Department of Materials Science and Nanoengineering, Rice University, Houston, Texas 77005, USA

[2]Department of Physics and Astronomy, University of Sussex, Brighton BN1 9RH, United Kingdom

[3]DEVCOM Army Research Laboratory, RF Devices and Circuits, Adelphi, Maryland 20783, USA

[4]Department of Chemistry, Rice University, Houston, Texas 77005, USA

[a)]**Authors to whom correspondence should be addressed:** abhijit.biswas@rice.edu, m.tripathi@sussex.ac.uk, biy@rice.edu, ajayan@rice.edu

**TABLE S1.** Comparative FWHM of h-BN thin films $E_{2g}$ Raman peak.

| h-BN thin film growth technique | Growth Morphology | $E_{2g}$ Raman peak FWHM (cm$^{-1}$) | References |
|---|---|---|---|
| CVD | Triangles (random orientation) | ~25 | [16] |
| CVD | Triangles (single orientation) | Not given | [17] |
| CVD | Oriented triangles | ~14.5 | [20] |
| CVD | Triangles (random orientation) | ~Not given | [24] |
| CVD | Triangles (random orientation) | ~ 20 | [26] |
| CVD | Triangles (random orientation) | ~Not given | [28] |
| CVD | Triangles (single orientation) | $\geq 25$ | [29] |
| MBE | Layered film | 14.1 | [34] |
| PLD | Amorphous | ~58 | [38] |
| PLD | Island growth | ~30.5 | [42] |
| **PLD** | **Triangles (single orientation)** | **~13-16** | **Our film** |
| Single crystal | Plate-like | ~8 | [37] |

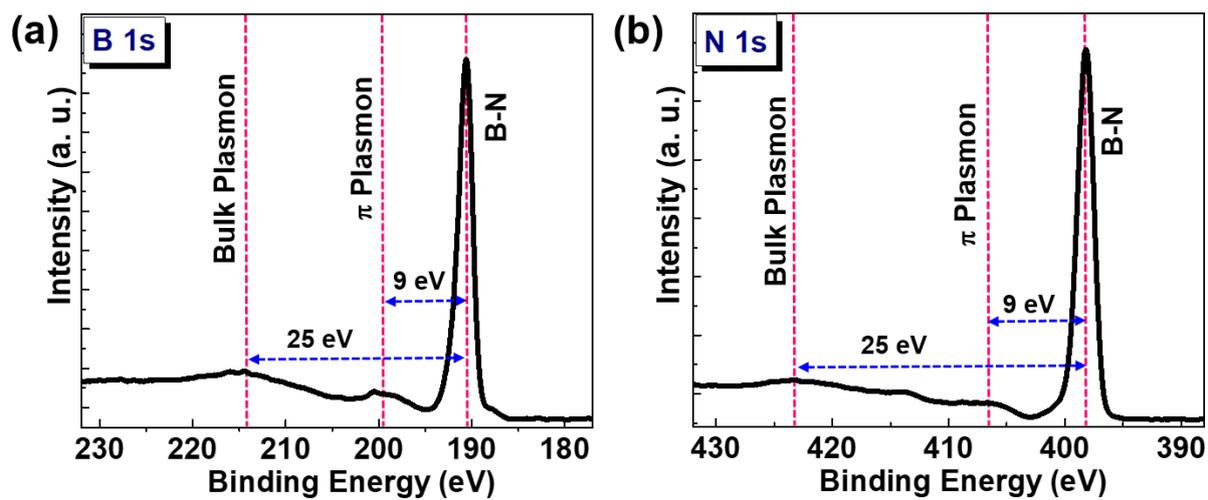

**FIGURE S1.** XPS spectra of a ~20 nm film show the characteristic B-N bonding as well as the π-Plasmon and bulk Plasmon peaks of h-BN **[32-34]**.

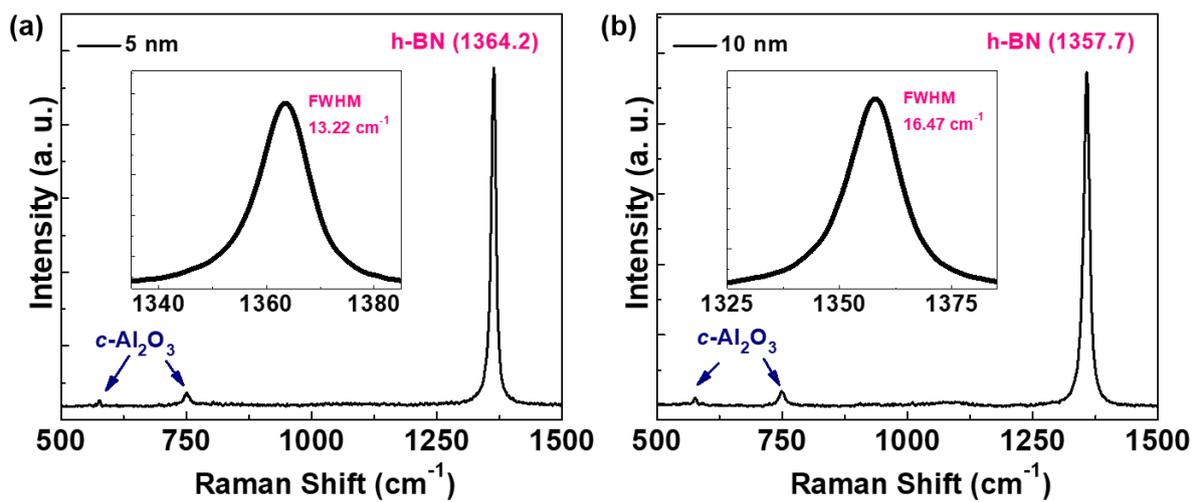

**FIGURE S2.** Raman spectra of films show the characteristic $E_{2g}$ Raman peak. Inset shows the full-width at half maxima (FWHM) of the observed Raman peak.

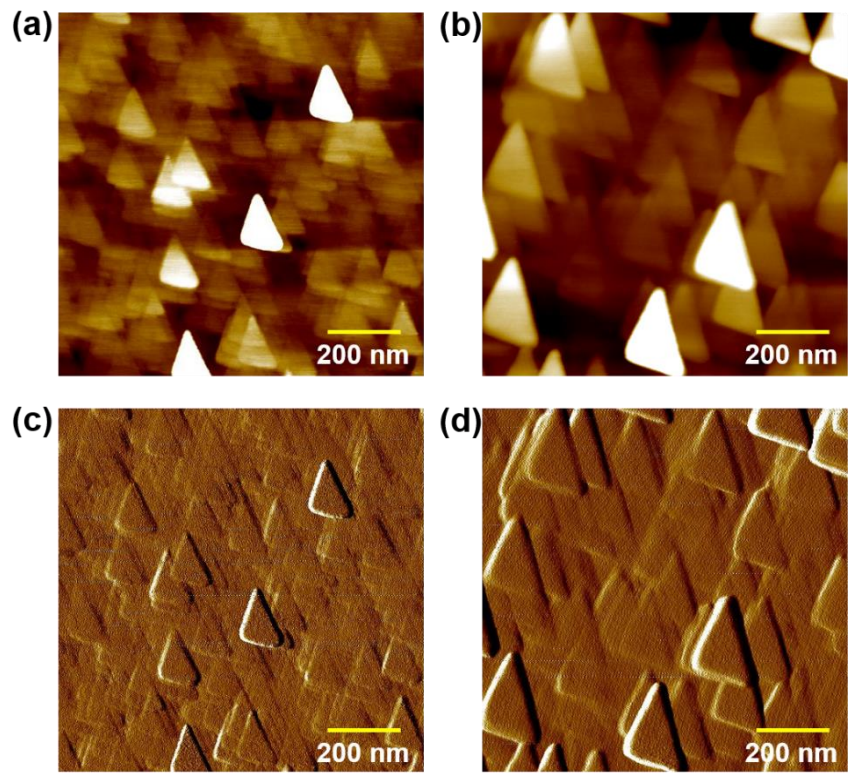

**FIGURE S3.** Atomic force microscopy images of various thicknesses h-BN film (~10 nm and ~20 nm). (a) Topography (upper panel) and (b) corresponding phase images (lower panel).

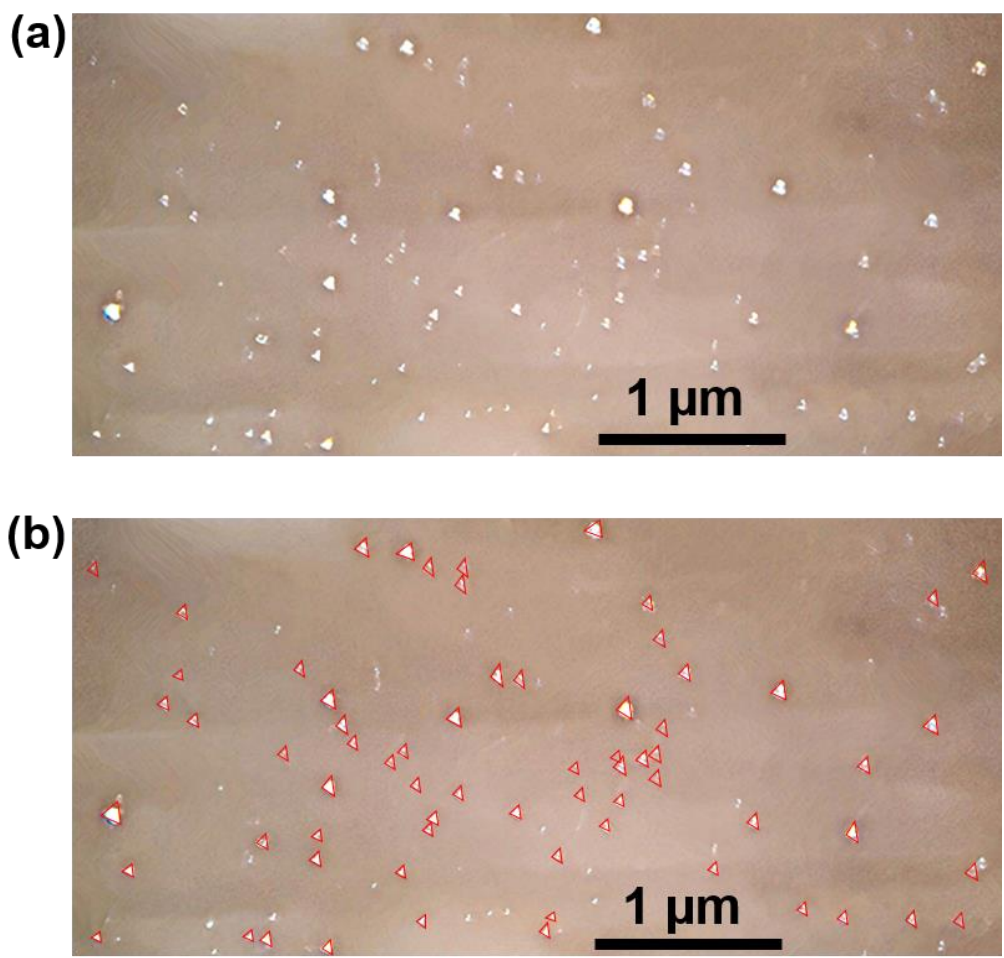

**FIGURE S4.** (a), (b) Optical microscope image (top: the actual image; bottom: same image with triangles are marked in red color) of a 5 nm BN film (500 laser shots) showing the unidirectional domains. The contrast of each individual domain is correlated to its thickness, i.e., thicker flakes appear much brighter (as similarly shown in AFM images).

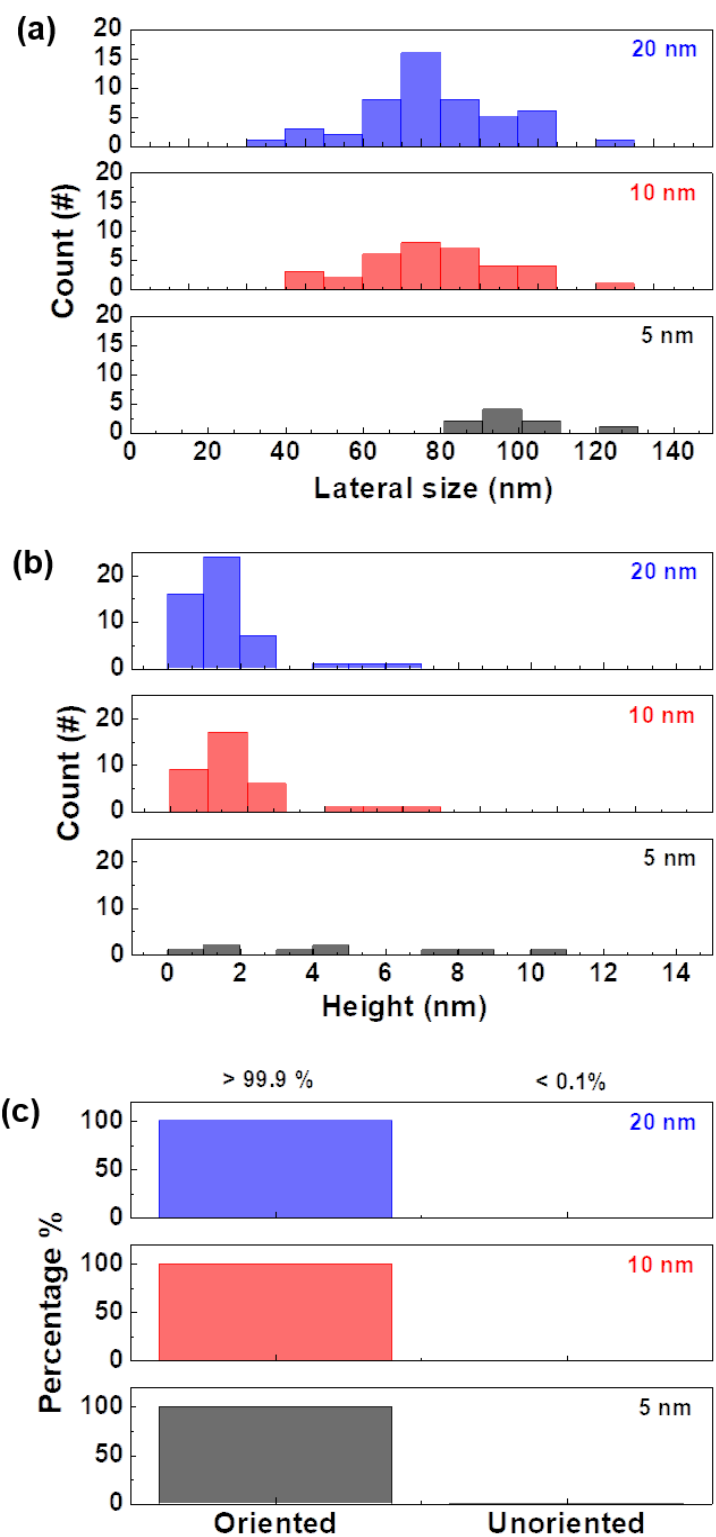

**FIGURE S5.** (a)-(c) Statistics regarding the lateral size, height (thickness) and orientation distribution of BN domains (5 nm, 10 nm and 20 nm corresponds to 500, 1000, and 2000 laser-shots film). The actual domain thickness distribution may not be accurate as for thicker films, domains are sitting on top of each other.

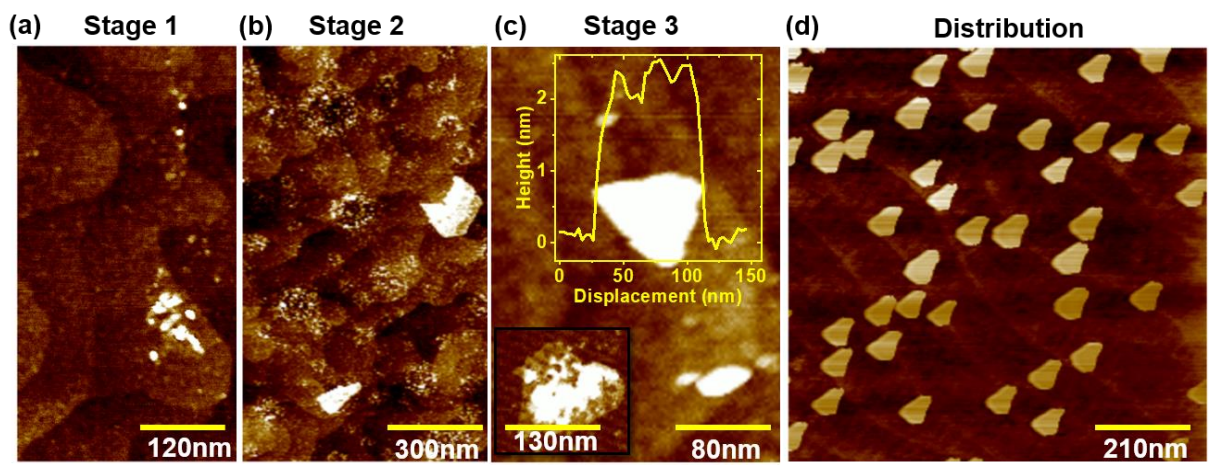

**FIGURE S6.** AFM morphology of nucleation, domain formation and step-edge guided growth. (a) Topography illustrating the initial stage "stage 1" of the accumulation of BN particles at the steps region (for only 5 laser shots film). (b) Topography showing the intermediate stage "stage 2" where merging of layers in the process to form triangular shape BN (for only 10-laser shots film). (c) Topography of a stage 3 showing the formed triangular domains (another region of the same laser 10 laser-shots film) (inset revealing formation the of a monolayer domain). (d) Phase image revealing the steps-edge guided growth where all the domains are at the step-edges (for100 laser-shots films).

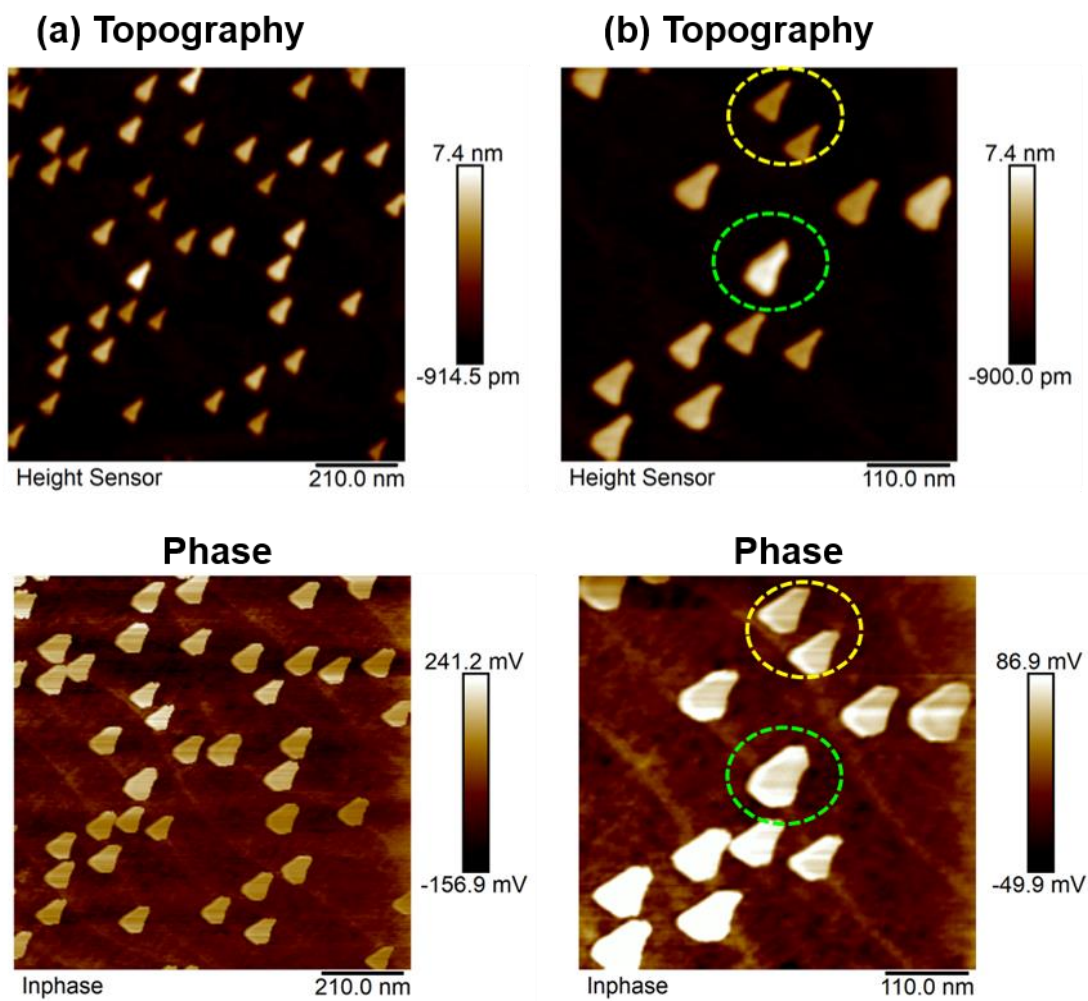

**FIGURE S7.** (a) Topography of different h-BN islands and their corresponding phase image. The topography image shows the triangular-shaped h-BN islands of minuscule distinguishable geometry. The phase contrast image reveals the step-edge beneath the h-BN islands as the nucleating sites. (b) The topography and phase image of the particular zoomed region show the difference in the geometry of h-BN marked in dashed circles indicating an artefact-free image.

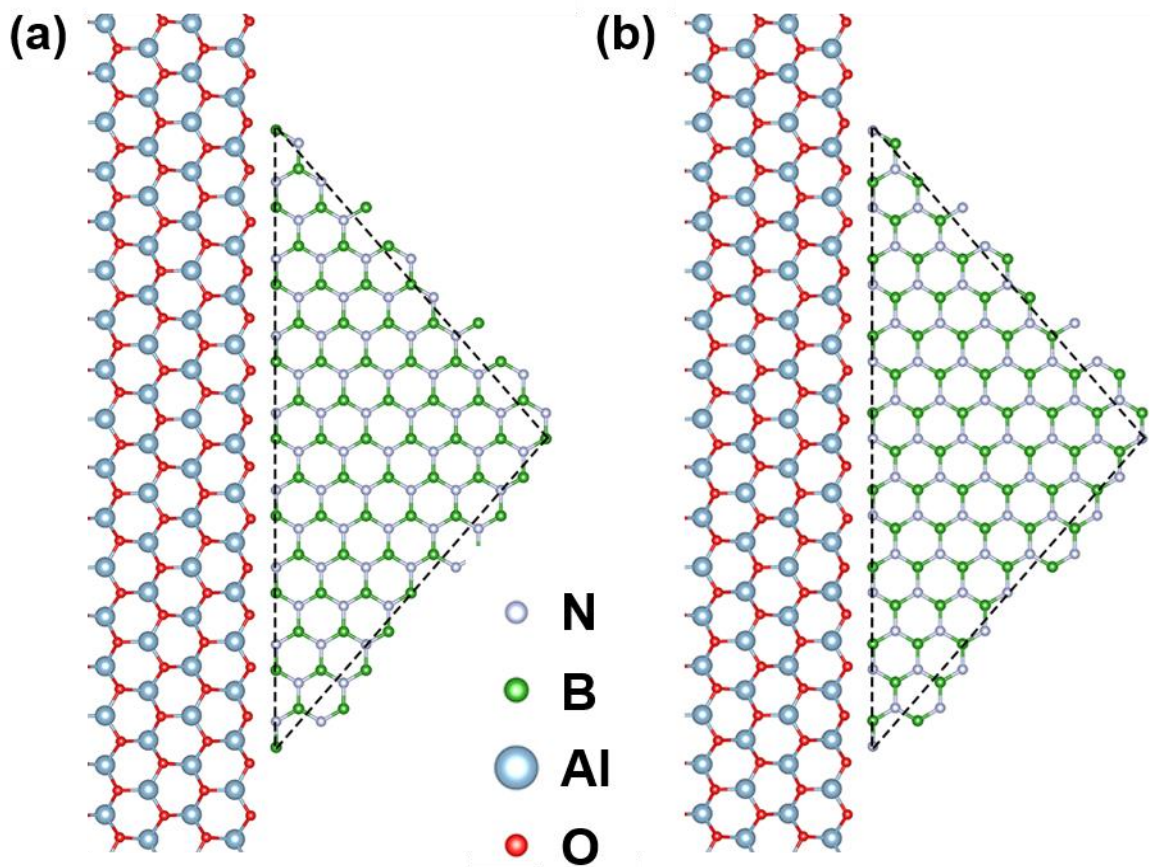

**FIGURE S8.** Top view of two different armchair (AR) h-BN edge attach to O-terminated sapphire (10-10) step-edge. (a) Unlike zigzag (ZZ) h-BN edge shown in the main text, AR edge will lead to two different edges, marked by dashed lines (some atoms outside the triangle are left for a better represent); therefore the two side angles of triangular domain will finally be different. (b) When swapping B/N atoms in (a), to *c*-$Al_2O_3$, the interface remains same, so does the interface binding energies of both (a) and (b), therefore, two types of islands will appear with the same probability.

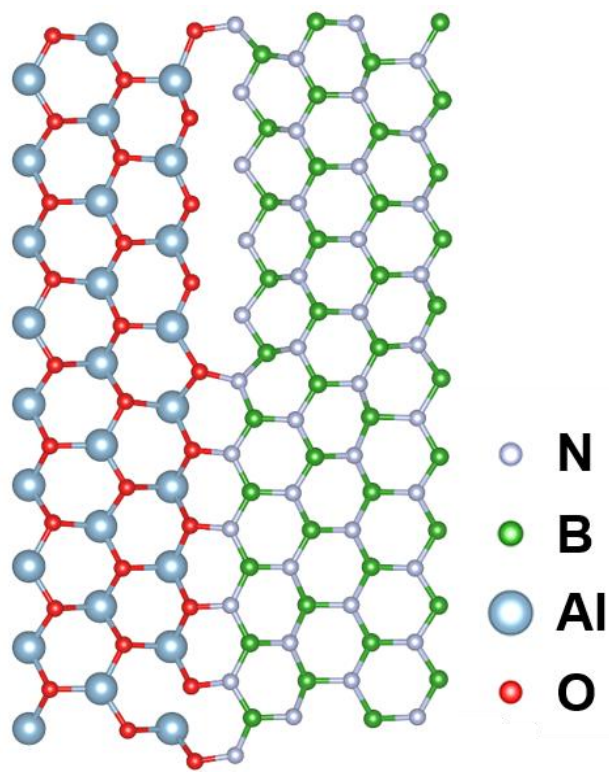

**FIGURE S9.** The relaxed structure of the interface between sapphire step-edge and ZZ-N terminated h-BN edge (only the Al/O atoms in the topmost layer are shown). The defect corresponds to the oxygen atoms leaving the plane and moves down close to the second $Al_2O_3$ layer.

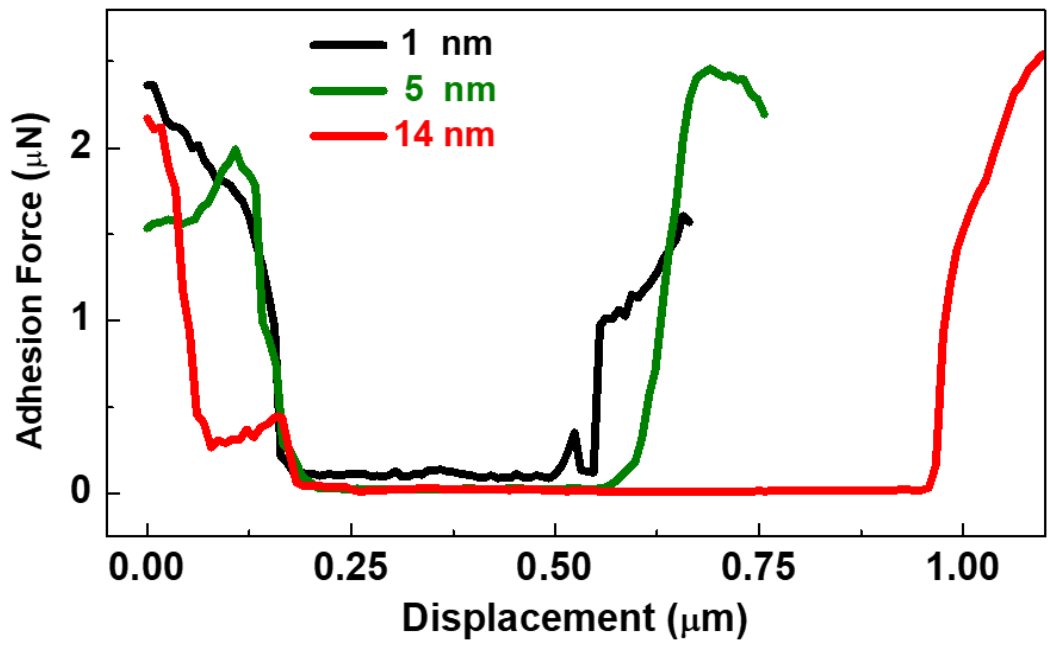

**FIGURE S10.** The adhesion force of h-BN film having various thicknesses.

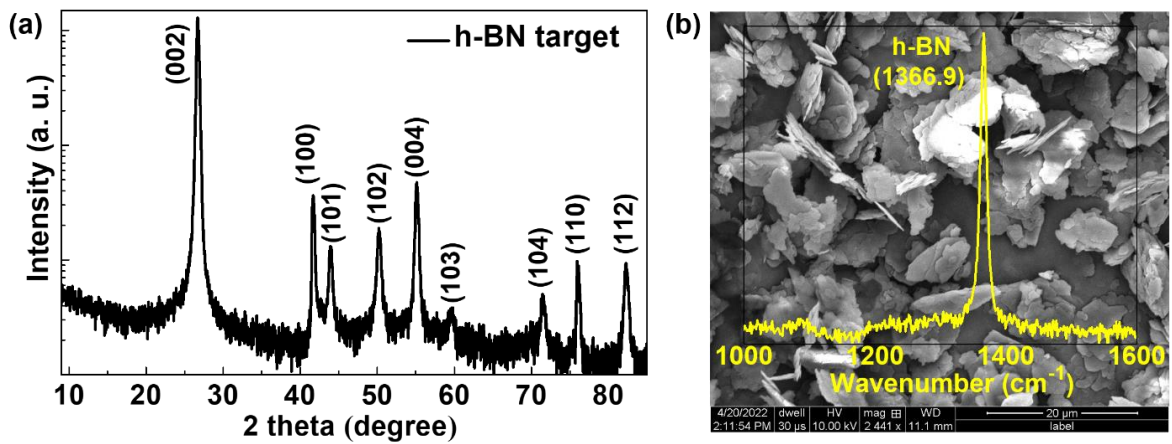

**FIGURE S11.** (a), (b) XRD, FESEM and Raman of the bulk h-BN polycrystalline target used for laser ablation.